\documentclass[aps,prl,twocolumn,showpacs]{revtex4}
\usepackage{graphicx} 
\usepackage{dcolumn}  
\usepackage{bm}       

\def\ii{{\rm i}}   \def\rb{{\bf r}}  \def\Eb{{\bf E}}
\def\ee{{\rm e}}   \def\Rb{{\bf R}}  \def\eh{{\epsilon_{\rm h}}}
\def\xx{\hat{\bf x}}   \def\Hb{{\bf H}}   
\def\yy{\hat{\bf y}}   \def\mb{{\bf m}}   \def\rp{{r_{\rm p}}}
\def\zz{\hat{\bf z}}   \def\pb{{\bf p}}

\begin{document}
\title{Electromagnetic surface states in structured perfect-conductor surfaces}
\author{F.~J.~Garc\'{\i}a~de~Abajo$^1$ and J.~J. S\'{a}enz$^2$}
\address{$^1$Centro Mixto CSIC-UPV/EHU and Donostia International Physics Center
(DIPC),
Apartado 1072, 20080 San Sebastian, Spain \\
$^2$Departamento de F\'{\i}sica de la Materia Condensada,
Universidad Aut\'{o}noma de Madrid, Cantoblanco, 28049 Madrid,
Spain}

\date{\today}

\begin{abstract}
Surface-bound modes in metamaterials forged by drilling periodic
hole arrays in perfect-conductor surfaces are investigated by
means of both analytical techniques and rigorous numerical
solution of Maxwell's equations. It is shown that these
metamaterials cannot be described in general by local,
frequency-dependent permittivities and permeabilities for small
periods compared to the wavelength, except in certain limiting
cases that are discussed in detail. New related metamaterials are
shown to exhibit exciting optical properties that are elucidated
in the light of our simple analytical approach.
\end{abstract}



\maketitle

Structured metal surfaces offer a playground to yield remarkable
optical phenomena ranging from extraordinary light transmission
through subwavelength hole arrays \cite{ELG98} to innovative types
of surface resonances \cite{PMG04}. The relevant role of surface
plasmons \cite{R1957} at visible and near-infrared frequencies was
emphasized in early developments \cite{ELG98,GBM04},
while subsequent studies, following pioneering works in the field
\cite{CE1961,M1980},
demonstrated similar effects in plasmon-free good conductor films
at microwave and THz frequencies \cite{GSH03,CN04}.
Recently, these two regimes have been connected by the exciting
prediction of Pendry {\em et al.} \cite{PMG04} of surface
resonances that mimic surface-plasmon behavior in
perfect-conductor surfaces (PCS's) textured with subwavelength
holes, and by its subsequent experimental observation
\cite{HES05}.

Surface plasmons, originally predicted by Ritchie \cite{R1957},
have given birth to the rapidly growing field of plasmonics owing
to their potential application in areas as diverse as biosensing
\cite{SSA93}, information processing via metal-surface circuits
\cite{BDE03}, or laser technology \cite{CST03}. Hence the
importance of devising new ways to achieve surface-plasmon-like
behavior in different frequency domains (e.g., using
phonon-polaritons in the infrared \cite{GCJ99,HTK02}, or textured
PCS's at lower frequencies \cite{PMG04,HES05}).

In this Letter, we introduce a new approach to systematically
study surface resonances in structured PCS's. This allows us to
provide further insight into recently proposed holey metamaterials
\cite{PMG04}, for which we find similar qualitative behavior and
significant quantitative corrections in the surface mode
dispersion relations. Furthermore, we show that this kind of
materials cannot be represented in general by local,
frequency-dependent optical constants [$\epsilon(\omega)$ and
$\mu(\omega)$], except in some limiting cases. Finally, our
results suggest a new systematics in the analysis of textured
PCS's that is applied to related metamaterial designs.

The material proposed by Pendry {\em et al.} \cite{PMG04} consists
of a planar PCS perforated by infinitely-long square holes of side
$a$ arranged in a periodic square array of period $d\ll\lambda$
and filled with homogeneous material of permittivity
$\epsilon_{\rm h}$ and permeability $\mu_{\rm h}$ ($\mu_{\rm h}=1$
will be used throughout this work), as sketched in the inset of
Fig.\ \ref{Fig1}a. The assumption that the field inside the
square-holes can be approximated by the lowest-frequency mode
(i.e., the TE$_{1,0}$ mode \cite{J1975}), allows them to describe
the bulk material by effective local optical constants
$\epsilon_\parallel$ and $\mu_\parallel$ for fields parallel to
the surface and $\epsilon_\perp=\mu_\perp=\infty$ for
perpendicular fields (PMG model \cite{PMG04}).
In such an effective medium, the reflection coefficient
for p-polarized incident light (external magnetic field parallel
to the surface) is given by Fresnel's equation
\begin{eqnarray}
\rp=\frac{k_z-k \sqrt{\mu_\parallel/\epsilon_\parallel}}{k_z+k
\sqrt{\mu_\parallel/\epsilon_\parallel}},
\label{Fresnel}
\end{eqnarray}
where $k=\omega/c$ is the free-space light momentum, and
$k_\parallel$ and $k_z=\sqrt{k^2-k_\parallel^2}$ are the momentum
components parallel and perpendicular to the surface,
respectively. The surface resonances are signalled by the
divergence of $\rp$ for incident evanescent waves
($k_\parallel>k$), leading to
\begin{eqnarray}
k_\parallel^2=k_{\rm SP}^2=k^2 + \Gamma \, \frac{A^3 k^4}{d^4}
\label{eqPendry}
\end{eqnarray}
with $A=a^2$ (the hole cross-section) and $\Gamma=64\mu_{\rm
h}^2/\pi^6$ in the long-wavelength limit \cite{com1}.

Fig.\ \ref{Fig1} illustrates through a representative example a
comparison between the PMG model \cite{PMG04} (dashed curves) and
a rigorous numerical solution of Maxwell's equations obtained by
expanding the electromagnetic field in terms of diffracted plane
waves outside the surface and square-waveguide modes inside the
holes. Our calculated results exhibit the same qualitative
behavior as the PMG model in the long-wavelength limit (Fig.\
\ref{Fig1}a), although, as expected, the inclusion of diffraction
orders outside the material introduces new structure in the
high-frequency region and bends the dispersion relation at the
boundary of the first Brillouin zone. Closer inspection into the
long-wavelength region reveals a sizable correction in the
position of the resonance towards smaller values of $k_\parallel$
(Fig.\ \ref{Fig1}b), quantified in a factor-of-3 larger decay
length of the surface state into the vacuum. The physical origin
of this discrepancy can be understood within the analytical
approach that follows, corroborated by full numerical solution of
Maxwell's equations.

We consider a unit p-polarized plane wave incident along the $xz$
plane (no surface modes are obtained for s polarization). The
material will be contained in the $z<0$ region. In the absence of
any surface structure, the total (incident plus reflected) field
reads
\begin{eqnarray}
\Eb^{\rm ext}(\rb)&=&\frac{2}{k} [\ii k_z \sin(k_z z) \xx -
k_\parallel \cos(k_z z) \zz]\,
\ee^{\ii k_\parallel x} \nonumber \\
\Hb^{\rm ext}(\rb)&=&2 \cos(k_z z) \yy\,\ee^{\ii k_\parallel x}.
\label{fields}
\end{eqnarray}
Now, the small hole limit ($a\ll d,\lambda$) allows representing
each hole by effective dipoles. This was shown by Bethe
\cite{B1944} for a single hole in a thin perfect-conductor screen,
where the field scattered from the hole is equivalent to that
generated by an electric dipole perpendicular to the surface plus
a parallel magnetic dipole, which are in turn proportional to the
external perpendicular electric field and parallel magnetic field
via the polarizabilities $\alpha_{\rm E}$ and $\alpha_{\rm M}$,
respectively. Parallel electric dipoles and perpendicular magnetic
dipoles are forbidden by the condition that the parallel electric
field and the perpendicular magnetic field vanish at a PCS. A
single hole in a semi-infinite PCS can be described in the same
fashion. In the electrostatic limit, $\alpha_{\rm E}$
($\alpha_{\rm M}$) can be calculated from the electrostatic
(magnetostatic) solution for an external electric (magnetic)
field, as shown in Fig.\ \ref{Fig2}a (Fig.\ \ref{Fig2}b), which
involves only TM modes (TE modes) of the hole cavity.

The actual self-consistent field acting on each hole includes
contributions from inter-hole dynamical interaction and symmetry
considerations show that the magnetic and electric dipoles must be
oriented as $\mb=m\yy$ and $\pb=p\zz$. Actually, $\mb$ and $\pb$
depend on hole positions $\Rb=(x,y)$ just through phase factors
$\exp(\ii k_\parallel x)$. This permits writing the
self-consistent relations \cite{CE1961}
\begin{eqnarray}
p&=&\alpha_{\rm E} [E_z^{\rm ext} + G_{zz}^{\rm EE} p + G_{zy}^{\rm EM} m] \nonumber \\
m&=&\alpha_{\rm M} [H_y^{\rm ext} + G_{yy}^{\rm MM} m +
G_{yz}^{\rm ME} p], \label{pandm}
\end{eqnarray}
where $G_{ij}^{\rm EE}=G_{ij}^{\rm MM}$ and $G_{ij}^{\rm
EM}=-G_{ij}^{\rm ME}=-G_{ji}^{\rm EM}$ are lattice sums of the
electric (E) and magnetic (M) dipole-dipole interactions, and $i$
and $j$ denote $y$ and $z$ Cartesian components. More precisely,
\begin{eqnarray}
G_{ij}^{\rm EE}&=&\sum_{{\bf R}\neq 0} \ee^{-\ii k_\parallel x}
(k^2+\partial_i\partial_j)\frac{{\rm e}^{{\rm i} kR}}{R} \nonumber\\
G_{yz}^{\rm EM}&=&-\ii k\sum_{{\bf R}\neq 0} \ee^{-\ii k_\parallel
x}\partial_x\frac{{\rm e}^{{\rm i} kR}}{R}. \label{GG}
\end{eqnarray}
This inter-hole interaction is generally small for $a\ll d$,
except when a diffraction order goes grazing, in which case the
above sums can diverge giving rise to phenomena related to Wood's
anomalies \cite{W1935}. It is near these divergences that
surface-bound modes can exist, subject to the condition
\begin{eqnarray}
(\alpha_{\rm E}^{-1}-G_{zz}^{\rm EE}) (\alpha_{\rm
M}^{-1}-G_{yy}^{\rm EE}) = (G_{yz}^{\rm EM})^2, \label{dispersion}
\end{eqnarray}
which is obtained from the vanishing of the secular determinant of
Eqs.\ (\ref{pandm}). In particular, near the light-line in the
$k_\parallel-\omega$ plane for $k_\parallel>k$, one has
\begin{eqnarray}
{\rm Re}\{G_{zz}^{\rm EE}\}\approx {\rm Re}\{G_{yy}^{\rm
EE}\}\approx {\rm Re}\{G_{yz}^{\rm EM}\}\approx \frac{2\pi\ii
k^2}{k_z d^2} \equiv S,
\end{eqnarray}
which comes from the divergent term of the parallel-momentum
expansion of Eqs.\ (\ref{GG}) (see Fig.\ \ref{Fig3}a).
Furthermore, upon inspection, one finds that ${\rm
Im}\{G_{yz}^{\rm EM}\}=0$, and the remaining imaginary parts of
all quantities in Eq.\ (\ref{dispersion}) cancel out exactly since
${\rm Im}\{G_{jj}^{\nu\nu}\}={\rm Im}\{\alpha_\nu^{-1}\}=-2k^3/3$
\cite{com2}, with $\nu=E,M$. Combining these results, Eq.\
(\ref{dispersion}) can be approximated by Eq.\ (\ref{eqPendry})
with
\begin{eqnarray}
\Gamma=\frac{4\pi^2}{A^3} (\frac{1}{{\rm Re}\{\alpha_{\rm
E}^{-1}\}}+\frac{1}{{\rm Re}\{\alpha_{\rm M}^{-1}\}})^2.
\label{gamma}
\end{eqnarray}
Eq.\ (\ref{gamma}) is exact in the $a\ll d\ll\lambda$ limit, and
it predicts the existence of surface-bound modes under the
condition $1/{\rm Re}\{\alpha_{\rm E}^{-1}\}+1/{\rm
Re}\{\alpha_{\rm E}^{-1}\}>0$. Calculated values of $\Gamma$ are
offered in Fig.\ \ref{Fig2}c for various systems. The position of
the surface mode calculated from Eq.\ (\ref{gamma}) (see Fig.\
\ref{Fig1}b) differs slightly from the exact numerical result,
mainly due to neighboring-holes multipolar interaction for
$a=0.8d$ (the holes occupy 64\% of the surface).

This description in terms of effective dipoles permits writing the
specular reflection coefficient for p-polarized light as $\rp=1+ S
(m-p k_\parallel/k)$. It is easy to see that this expression does
not conform in general to the assumption of local optical
constants implicit in the derivation of Eq.\ (\ref{Fresnel}). In
particular, for non-grazing incidence and $\lambda\gg d$, the
dipole-dipole interaction can be neglected in Eqs.\ (\ref{pandm}),
so that using explicit expressions for the fields as provided by
Eqs.\ (\ref{fields}), and noticing the $\rp$ deviates only
slightly from unity under these conditions, one finds
\begin{eqnarray}
\sqrt{\mu_\parallel/\epsilon_\parallel}\approx\frac{2\pi\ii
k}{d^2}(\alpha_{\rm M}+\alpha_{\rm E} (k_\parallel/k)^2),
\label{meta}
\end{eqnarray}
which is independent of $k_\parallel$ (i.e., of the angle of
incidence) only if $\alpha_{\rm E}=0$. Otherwise, the metamaterial
is non-local, so that the optical constants of an equivalent
homogeneous medium will depend on both frequency and momentum
(spatial dispersion). It should be noted that the neglect of
cavity modes other than the lowest-frequency one (TE$_{1,0}$)
yields $\alpha_{\rm E}=0$ (see Fig.\ \ref{Fig2}c), and therefore,
it leads to an incomplete local-response description of the
metamaterial \cite{PMG04}.

From the point of view of external fields, the local-response
picture will be still maintained if $|\alpha_{\rm
E}|\ll|\alpha_{\rm M}|$, so that the second term in the right hand
side of Eq.\ (\ref{meta}) can be overlooked. Such metamaterials
can be achieved by filling the holes with media of very-high
$|\eh|\gg 1$ or alternatively by using specific electrostatic
resonances in the $-1<\eh<0$ range (piling up towards $\eh=-1$ as
a manifestation of their filling-material surface-plasmon origin).
This makes $\alpha_{\rm E}$ negligible in the long-wavelength
limit, as shown in Fig.\ \ref{Fig4}.

Another possibility consists in filling dimples rather than holes
using moderate values of $\eh>1$ (e.g., $\eh=10$ in Fig.\
\ref{Fig5}) and $\lambda/d$. Indeed, the high-frequency
propagating modes of infinitely-long holes become resonances of
finite width (coupling to the external continuum) in holes of
finite depth (dimples), which are blue shifted with respect to the
noted propagating modes owing to reflection at the bottom and top
ends of the hole (Fabry-Perot picture), as illustrated by
comparing Fig.\ \ref{Fig3}b and Fig.\ \ref{Fig3}c. For the choice
of parameters of Fig.\ \ref{Fig3}c, there is a resonance in
$\alpha_{\rm M}$ near the lowest-frequency cavity mode ($\lambda=
7.2 a$), where $\alpha_{\rm E}$ is comparatively negligible. This
dimple resonance provides a cut-off of surface-bound modes
\cite{PMG04}, as illustrated in Fig.\ \ref{Fig5}. Higher-energy
resonances of both TM and TE nature are also observed in the
reflectivity, giving rise to a rich structure of surface-bound
states (Fig.\ \ref{Fig5}).

Similar behavior could be obtained by exploiting the mentioned
single-hole electrostatic resonances. Optical phonons in alkali
halides yield $\eh<0$ \cite{com3} and can be combined with noble
metals (near-perfect conductors) to implement these ideas in the
THz domain. From Eqs.\ (\ref{eqPendry}) and (\ref{gamma}) and from
the electric polarizability given in Fig.\ \ref{Fig4}, the
frequency dependence of $\eh(\omega)$ produces dispersion
relations similar to Fig.\ \ref{Fig5} (not shown), including a
cut-off due to the electrostatic resonances near $\eh=-1$.

The surface modes resemble surface plasmons not only in their
limited penetration into the vacuum, but also in the interaction
that they provide between additional surface features like holes
of larger dimensions. Indeed, the scattered field produced by one
of such features in the metamaterial decays away along the surface
as $\exp(\ii k_{\rm SP} R)/\sqrt{R}$ at large distance $R$. This
has the same form as the charge distribution accompanying a
surface plasmon launched by a localized source \cite{KOP05}, in
contrast to the $\exp(\ii k R)/R$ far-field dependence of the
interaction on unstructured PCS's \cite{com4}. Furthermore, the
far field of a small, localized additional surface feature can be
assimilated to the field of an effective dipole placed at the
surface of an equivalent homogeneous material with the same
reflectivity as the holey metamaterial. Interestingly, in contrast
to the dipoles that describe the underlying hole structure, the
new effective dipole can have parallel electric and perpendicular
magnetic components. This introduces another handle in the design
of surface states by using the above holey metamaterials as the
base fabric to build metamaterials drilled by larger holes,
allowing us to speculate on surface modes in fractal structures
that imitate Sierpinski's carpet.


In conclusion, we have introduced a formalism to study textured
perfect-conductor surfaces that allows us to obtain
quasi-analytical long-wavelength exact dispersion relations for
surface-bound modes. We have found that these metamaterials cannot
be assimilated in general to equivalent effective homogeneous
media described by local optical constants, except in some
limiting cases (e.g., by filling the holes with
high-index-of-refraction material). Finally, our results pave the
way towards simple analysis of new metamaterial designs based upon
the coexistence of different hole sizes and distributions that can
realize the goal of achieving tailored surface dispersion
relations.

\begin{acknowledgments}
FJGA would like to thank Prof. Pendry for helpful suggestions and
for very enjoyable and stimulating discussions. This work was
supported in part by the Spanish MEC (FIS2004-06490-C03-02 and
BFM2003-01167) and by the European Commission ({\em Metamorphose}
NoE NMP3-CT-2004-500252 and {\em Molecular Imaging} IP
LSHG-CT-2003-503259).
\end{acknowledgments}


\begin{figure}
\caption{\label{Fig1} {\bf (a)} Modulus of the specular reflection
coefficient $|\rp|$ of a perfect-conductor surface (PCS)
perforated by infinitely-long square holes (see inset for
parameters) as a function of wavelength $\lambda$ and momentum
parallel to the surface $k_\parallel$. The surface mode predicted
in Ref.\ \onlinecite{PMG04} is shown by a dashed curve. {\bf (b)}
Detail of the reflection coefficient $\rp$ along the AB segment of
(a).}
\end{figure}

\begin{figure}
\caption{\label{Fig2} {\bf (a)} Electrostatic electric-field flow
lines for a hole drilled in a semi-infinite perfect-conductor
subject to an external field $\Eb^{\rm ext}$ perpendicular to the
surface, giving rise to an electric dipole $p=\alpha_{\rm E}
E^{\rm ext}$ as seen from afar. {\bf (b)} Magnetostatic
magnetic-field flow lines for the same hole subject to an external
parallel field $\Hb^{\rm ext}$ and leading to a magnetic dipole
$m=\alpha_{\rm M} H^{\rm ext}$. {\bf (c)} Summary of
polarizabilities for square and circular holes in PCS's,
normalized using the hole area $A$. The circular hole in a
perfect-conductor thin screen is analytical
\cite{B1944,J1975,com5}.}
\end{figure}

\begin{figure}
\caption{\label{Fig3} {\bf (a)} Dipole-dipole interaction sums for
a square lattice of period $d$ [Eqs.\ (\ref{GG}) for
$k_\parallel\gtrsim k$]. {\bf (b)} Electric (solid curve) and
magnetic (broken curve) polarizabilities of single holes filled
with $\eh=10$ material as a function of hole size. {\bf (c)} Same
as (b) for dimples.}
\end{figure}

\begin{figure}
\caption{\label{Fig4} Inverse of the electrostatic polarizability
of a single circular hole of area $A$ as a function of the
dielectric constant $\eh$ of the filling material. The field
strength distribution is shown in the insets for the two
lowest-order resonances in the $-1<\eh<0$ range. The magnetostatic
response is independent of $\eh$ ($A^{3/2}/\alpha_{\rm M}\approx
37.3$).}
\end{figure}

\begin{figure}
\caption{\label{Fig5} Same as Fig.\ \ref{Fig1}a for holes of
finite depth $t=1.5 a$ filled with material of dielectric constant
$\eh=10$ (see inset).}
\end{figure}


\begin{thebibliography}{25}
\bibitem[{com({\natexlab{a}})}]{com1}
\bibinfo{note}{This value of $\Gamma$ is obtained assuming that $\omega$ is
  small compared to $\omega_{\rm pl}$ in Eq.\ (12) of Ref.\
  \onlinecite{PMG04}.}
\bibitem[{com({\natexlab{b}})}]{com2}
\bibinfo{note}{This exact, optical-theorem-like result is derived from flux
  conservation upon arbitrary external illumination for $k_\parallel>k$.}
\bibitem[{com({\natexlab{e}})}]{com5}
\bibinfo{note}{Notice that Eq.\ (\ref{gamma}) must be corrected by a factor of
  4 in thin screens due to cooperative interaction between both sides of the
  film.}
\end{thebibliography}

\end{document}